# Encryption in ghost imaging with Kronecker products of random matrices

Yi-Ning Zhao, Lin-Shan Chen, Lingxin Kong, Chong Wang, Cheng Ren, and De-Zhong Cao*

*Department of Physics, Yantai University, Yantai 264005, China*
*Corresponding author: dzcao@ytu.edu.cn*

By forming measurement matrices with the Kronecker product of two random matrices, image encryption in computational ghost imaging is investigated. The two-dimensional images are conveniently reconstructed with the pseudo-inverse matrices of the two random matrices. To suppress the noise, the method of truncated singular value decomposition can be applied to either or both of the two pseudo-inverse matrices. Further, our proposal facilitates for image encryption since more matrices can be involved in forming the measurement matrix. Two permutation matrices are inserted into the matrix sequence. The image information can only be reconstructed with the correct permutation matrices and the matrix sequence in image decryption. The experimental results show the facilitations our proposal. The technique paves the way for the practicality and flexibility of computational ghost imaging.

1. Introduction

In recent years, the rapid developments of ghost imaging (GI) have attracted much more attention of the researchers. Ghost imaging is an imaging method relied on the correlation of the measurement matrix and bucket detection signals. How to improve the efficiency of the acquisition and reconstruction is always the focus in the field of GI. Computational ghost imaging (CGI) [1] simplifies the setup of the GI experiments. As a result, the process of acquisition has been efficiently optimized in CGI. Measurement matrices with specific constructions have also been explored to improve the efficiency of image reconstruction in the CGI experiments. Recently, various kinds of orthogonal transform bases have been used to generate the measurement matrix. Fourier [2,3], discrete cosine [4-6], Hadamard [7-10], and wavelet [11] transform matrices have already been used as the measurement matrix in CGI. Images with high qualities can be reconstructed by the full use of unitarities of these matrices. Hybridization advantages of CGI are explored by hybridizing different orthogonal transform matrices [12]. The measurement matrix in experiment is formed by the Kronecker product of two different orthogonal transform matrices.

Indeed, the most classic kind of measurement matrix in GI is a random matrix, which is composed of random speckles. Classical sources [13] include true thermal light [14], pseudo-thermal light [15], sunlight [16], terahertz wave [17], X ray [18,19], and etc. [20,21] have been applied in experimental investigations. Specifically, the elements in random matrices satisfy a certain probability distribution, which also determines the statistics of the signals of the bucket detector [22]. The reconstruction efficiency of CGI with random measurement matrix needs to be further improved. Pseudo-inverse ghost imaging [23] and Schmidt ghost imaging [24,25] have been proposed to deal with the random measurement matrix. However, when facing the measurement matrices of big size, it is complex to calculate the pseudo-inverse matrices or perform Schmidt orthogonalization. And the time cost of them is really high, which resists the reconstruction efficiency of CGI by random matrices. Hence, how to use matrix hybridization to improve CGI with random speckles is still a challenge.

In this paper, we propose a scenario of CGI, in which the measurement matrix is generated by the Kronecker product of two smaller random matrices. The two-dimensional images can be high-efficiently reconstructed with the pseudo-inverse matrices of the two random matrices. In the reconstruction process of the image, truncated singular value decomposition (TSVD) is applied to either or both the two random matrices to suppress the noise. Experiments are performed with a real object, which contains two handmade windmills. The reconstruction efficiency of ghost imaging with Kronecker product of random matrices is much higher than traditional GI with random speckles.

Nowadays, ghost imaging has also been developed in the areas of image processing such as encryption. Various methods have been used to realize the encryption combined with ghost imaging. QR code [26], XOR operation [27], deep learning [28], and Arnold transformation [29] et al. have been used in the image encryption based on the ghost imaging. In addition, ghost imaging with Kronecker product of random matrices takes the advantages of matrix arrangement to demonstrate flexibility of image encryption. By inserting permutation matrices into the matrix sequence, encryption is realized. The bucket detection signals are disordered and encrypted. Only if the random matrices, permutation matrices, and their sequence information are known and applied correctly, the image decryption can be successful. Ghost imaging with

Kronecker product of random matrices greatly extends flexible applications in image encryption.

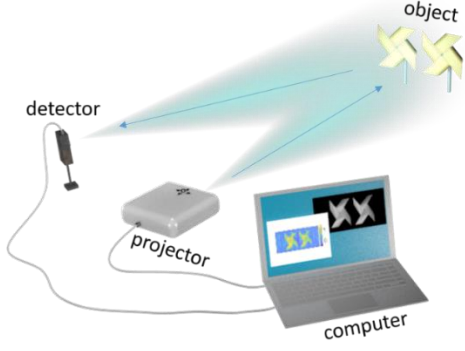

Fig. 1. Experimental setup.

2. Theory

In CGI, a series of random patterns are projected onto the object. A bucket detector is used to collect the object beam. In general, the reshaped random patterns play the role as the row vectors of the measurement matrix. The bucket detection signals are

$$y_{M\times 1} = A_{M\times N} x_{N\times 1}, \quad (1)$$

where $x_{N\times 1}$ is the object in one-dimensional form, and $A_{M\times N}$ is the measurement matrix composed of $M$ row vectors. The object information can be reconstructed through the intensity correlation function between the bucket detection signals and the measurement matrix that

$$x'_{N\times 1} = A^T_{N\times M} y_{M\times 1}, \quad (2)$$

where $T$ represents matrix transpose. For simplicity of representation, the footnotes of matrix sizes are ignored below.

The image reconstructed in ghost imaging often contains a constant background [30]. And the image visibility is often less than 1/3. In early investigations of ghost imaging with thermal light, some efforts, such as differential method [31], high-order correlation function [32], and compressed sensing [33], were made to improve the quality of reconstructed images.

We have proposed a scenario of ghost imaging with hybrid transforms by the Kronecker product of orthogonal matrices [12], and the imaging scheme can be written as

$$LXR^T = Y, \quad (3)$$

where $X$ is the object in two-dimensional form, $L$ and $R$ are the left and right transform matrices, $Y$ is the bucket detection signals in two-dimensional form. In eq. (3), $L$ and $R$ can be of different sizes, and from different transform matrices.

Due to the unitarity of matrices $L$ and $R$, the reconstructed image can be represented as

$$X = L^T Y R, \quad (4)$$

and the measurement matrix is

$$A = L \otimes R, \quad (5)$$

where $\otimes$ represents the Kronecker product.

In the case of ghost imaging with Kronecker product of random matrices, random matrices replace unitary matrices in the above Eqs. (3), (4), and (5). Similarly, the measurement matrix is obtained from the Kronecker product of the two random matrices as in Eq. (5). And the reconstructed image can be obtained from

$$X = L^{-1} Y (R^T)^{-1}. \quad (6)$$

In general, the inverse matrices of $L$ and $R^T$ are hard to find, due to their randomness and singularity.

However, the image reconstruction in Eq. (6) can also be implemented, just by replacing the inverse matrices with pseudo-inverse ones, as in pseudo-inverse ghost imaging [23]. Now the point is to find the pseudo-inverse matrices of random matrices $L$ and $R$, according to Moore-Penrose theorem [34].

Singular value decomposition (SVD) is a matrix factorization method, which is a concept in linear algebra. SVD can represent a complex matrix by the multiplication of several simpler sub-matrices, which describe the important properties of the matrix. According to singular value decomposition, the random matrices $L$ and $R$ are decomposed as

$$L = U_L S_L V_L^T, R = U_R S_R V_R^T, \quad (7)$$

where $U$ and $V$ are unitary matrices, and $S$ is a diagonal matrix. Then the measurement matrix becomes

$$A = (U_L \otimes U_R)(S_L \otimes S_R)(V_L^T \otimes V_R^T) = U_A S_A V_A^T, \quad (8)$$

where $U_A = U_L \otimes U_R, S_A = S_L \otimes S_R, V_A^T = V_L^T \otimes V_R^T$.

From Moore-Penrose theorem [34], the pseudo-inverse matrix of $A$ is

$$\widetilde{A}^{-1} = (V_L S_L^{-1} U_L^T) \otimes (V_R S_R^{-1} U_R^T) = \widetilde{L}^{-1} \otimes \widetilde{R}^{-1}, \quad (9)$$

where $\widetilde{L}^{-1} = V_L S_L^{-1} U_L^T$ and $\widetilde{R}^{-1} = V_R S_R^{-1} U_R^T$ are the pseudo-inverse matrices of the random matrices $L$ and $R$, respectively.

Substituting the pseudo-inverse matrices above into Eq. (6), an approximate image can be reconstructed by

$$\widetilde{X} \simeq \widetilde{L}^{-1} Y (\widetilde{R}^T)^{-1}, \quad (10)$$

where $(\widetilde{R}^T)^{-1} = (\widetilde{R}^{-1})^T$.

The TSVD method is efficient to relieve noise in GI [35]. By truncating the diagonal matrices $S_L$ and $S_R$ in Eq. (7), the random matrices $L$ and $R$ can be approximately represented as

$$L \simeq U_L^{(r_1)} S_L^{(r_1)} (V_L^T)^{(r_1)}, R \simeq U_R^{(r_2)} S_R^{(r_2)} (V_R^T)^{(r_2)}, \quad (11)$$

where $r_1$ and $r_2$ are the truncated rates of matrices $S_L$ and $S_R$, respectively, $U^{(r)}, S^{(r)}, (V^T)^{(r)}$ represent the truncated matrices $U, S, V^T$, respectively.

According to Eq. (11), the pseudo-inverse matrices of the random matrices $L$ and $R$ can be approximately represented by $\widetilde{L}^{-1} \simeq V_L^{(r_1)} (S_L^{(r_1)})^{-1} (U_L^{(r_1)})^T$ and $\widetilde{R}^{-1} \simeq V_R^{(r_2)} (S_R^{(r_2)})^{-1} (U_R^{(r_2)})^T$. Applying them into Eq. (10), the image of object can be reconstructed with less noise.

We can see that the truncated singular value decomposition of $\widetilde{A}^{-1}$ can be implemented by either or both of the truncated singular value decompositions of $\widetilde{L}^{-1}$ and $(\widetilde{R}^T)^{-1}$.

## 3. Experiment

The experimental setup is shown in Fig. 1. A projector (LG: PF1500G-GL) is used to project the illumination patterns on the object, and a charge coupled device (IMAVISION: MER-132-43GC-P) is used to collect the bucket detection signals. A computer (CPU: Intel Core i7-11700K, RAM: 128GB) is used to generate the matrices of the illumination patterns which form the measurement matrix, and finally reconstruct the image of the object.

Experiments of ghost imaging with Kronecker product of random matrices are performed with a real object, which are two handmade windmills as shown in Fig. 1 and in Fig. 2(a). We prepare two random binary (0 and 1) matrices $L = N_{32\times32}$ and $R = N_{64\times64}$ with probability of 0.2, as shown in Fig. 2(c). And the sampling number in the experiment is $32 \times 64 = 2{,}048$. The TSVD method is adopted by applying truncation rate 0.9 to both matrices $L = N_{32\times32}$ and $R = N_{64\times64}$. The reconstructed image with $32 \times 64 = 2{,}048$ pixels is shown in Fig. 2(b). As we can see, the quality of the reconstructed image is quite good. It is not surprising since the method of TSVD can suppress the noise and improve image quality. Peak signal-to-noise ratio (PSNR) and structural similarity (SSIM) [36] are used to estimate the quality of the reconstructed image. Because the object is a gray real object, it is not so easy to calculate the PSNR and SSIM of the reconstructed image directly. In this paper, an empty area of the object in the red frame is selected to estimate the quality of the reconstructed image. The PSNR and SSIM of the image in the red frame in Fig. 2(b) are $10.71\,dB$ and $2.59 \times 10^{-5}$, respectively. The SSIM value is very small since the area of the object in the red frame is empty.

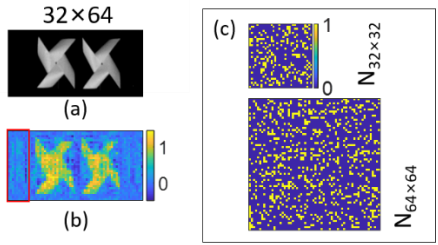

Fig. 2. Experimental results of ghost imaging with Kronecker product of random matrices with a real object. (a) The object (handmade windmills) with resolution of $32 \times 64$. (b) The reconstructed image. (c) The random binary matrices $L = N_{32\times32}$ (up) and $R = N_{64\times64}$ (down).

The total time taken to calculate the pseudo-inverse matrices $\widetilde{L}^{-1}$ and $(\widetilde{R}^T)^{-1}$ is $8.02 \times 10^{-4}\,s$, and the time of image reconstruction by $\widetilde{L}^{-1} Y (\widetilde{R}^T)^{-1}$ is $1.07 \times 10^{-5}\,s$. While the time taken to directly calculate the pseudo-inverse matrix $\widetilde{A}^{-1}$ (without using the Kronecker product) is $1.54\,s$, the time of image reconstruction by $\widetilde{A}^{-1} y$ is $1.06 \times 10^{-3}\,s$. It can be seen that under the circumstance of image with $32 \times 64$ pixels, the total time taken to calculate the pseudo-inverse matrices $\widetilde{L}^{-1}$ and $(\widetilde{R}^T)^{-1}$ is 0.05% of the time taken to calculate the pseudo-inverse matrix $\widetilde{A}^{-1}$, and the time taken to reconstruct the image by $\widetilde{L}^{-1} Y (\widetilde{R}^T)^{-1}$ is 1% of the time taken to reconstruct the image by $\widetilde{A}^{-1} y$. Ghost imaging with Kronecker product of random matrices has realized the improvement of the efficiency in image reconstruction and time cost is reduced.

## 4. Image encryption

A unitary permutation matrix $P$ was applied to the bucket detection signals to perform image encryption by Zhang et al. [37]. The permutation matrix $P$ has only one of 1 in each row and column, and others are 0. Its inverse matrix is its transpose. In this paper, we adopt two permutation matrices $P_1$ and $P_2$ which are placed in the left and right side of the image $X$ in Eq. (3). And then we can get 4 forms to realize the encryption, which are

$$P_1 L X P_2 R^T = Y_1,$$
$$P_1 L X R^T P_2 = Y_2,$$
$$L P_1 X P_2 R^T = Y_3, \quad (12)$$
$$L P_1 X R^T P_2 = Y_4.$$

The corresponding measurement matrices of Eq. (12) are

$$A_1 = (P_1 L) \otimes (R P_2^T),$$
$$A_2 = (P_1 L) \otimes (P_2^T R),$$
$$A_3 = (L P_1) \otimes (R P_2^T), \quad (13)$$
$$A_4 = (L P_1) \otimes (P_2^T R).$$

The reconstruction process can be written as

$$X_1 = L^{-1} P_1^T Y_1 (R^T)^{-1} P_2^T,$$
$$X_2 = L^{-1} P_1^T Y_2 P_2^T (R^T)^{-1},$$
$$X_3 = P_1^T L^{-1} Y_3 (R^T)^{-1} P_2^T, \quad (14)$$
$$X_4 = P_1^T L^{-1} Y_4 P_2^T (R^T)^{-1}.$$

Only the proper inverses of the permutation matrices and their positions are adopted in the reconstruction process, the correct image can be reconstructed. Others, image reconstruction will fail.

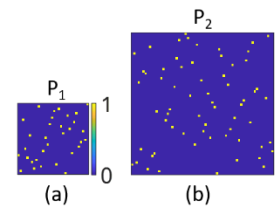

Fig. 3. The two permutation matrices. (a) The permutation matrix $P_1$ with size of $32 \times 32$. (b) The permutation matrix $P_2$ with size of $64 \times 64$.

Experiments of image encryption are performed with the object handmade windmills, too. The two permutation matrices used in the experiments are shown in Fig. 3. The two random matrices $L = N_{32\times32}$ and $R = N_{64\times64}$, which are shown in Fig. 2(c), are still used in the encryption experiments. The results of image encryption in ghost imaging with Kronecker product of random matrices are shown in Fig. 4. The images in the first (a1, b1, c1, d1), second (a2, b2, c2, d2), third (a3, b3, c3, d3), fourth (a4, b4, c4, d4), and fifth (a5, b5, c5, d5) column are reconstructed with reconstruction methods $\widetilde{N}_1^{-1}Y(\widetilde{N}_2^T)^{-1}$, $\widetilde{N}_1^{-1}P_1^T Y(\widetilde{N}_2^T)^{-1}P_2^T$, $\widetilde{N}_1^{-1}P_1^T Y P_2^T(\widetilde{N}_2^T)^{-1}$, $P_1^T\widetilde{N}_1^{-1}Y(\widetilde{N}_2^T)^{-1}P_2^T$, $P_1^T\widetilde{N}_1^{-1}Y P_2^T(\widetilde{N}_2^T)^{-1}$, respectively. The images in the Figs. 4(a2), 4(b3), 4(c4), and 4(d5) are successfully decrypted with proper sequence information of the two permutation matrices.

From the experimental results in Fig. 4, we can see that the permutation matrices and their positions in the matrix sequence are both important in image encryption. Once the image is encrypted by ghost imaging with Kronecker product of random matrices, image decryption will be more difficult than that in conventional GI. The PSNRs and SSIMs of the image in the red frame in Fig. 4 are calculated to estimate the qualities of the decrypted images. The PSNRs and SSIMs of the images in Figs. 4(a2), 4(b3), 4(c4), and 4(d5) are much greater than that in other groups, because they are decrypted successfully.

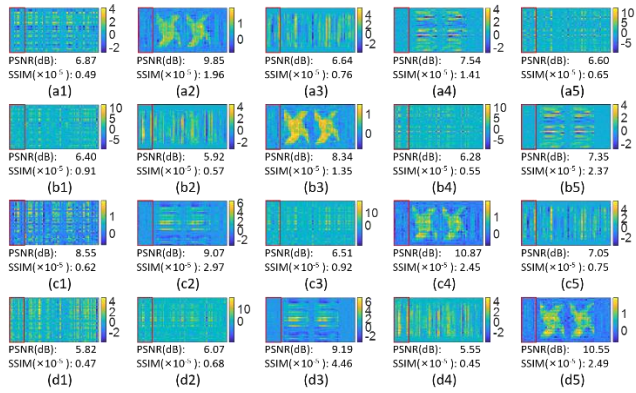

Fig. 4. Results of image encryption in ghost imaging with Kronecker product of random matrices. The images in the first (a1, a2, a3, a4, a5), second (b1, b2, b3, b4, b5), third (c1, c2, c3, c4, c5), and fourth (d1, d2, d3, d4, d5) row are reconstructed with bucket detection signals $P_1N_1XP_2N_2^T$, $P_1N_1XN_2^TP_2$, $N_1P_1XP_2N_2^T$, and $N_1P_1XN_2^TP_2$, respectively.

## 5. Conclusion

CGI with Kronecker product of random matrices is proposed in this paper. The Kronecker product of two random matrices is used as the measurement matrix. The method of TSVD, which is applied to the two submatrices, is used to calculate the pseudo-inverse matrices of the random matrices during the image reconstruction. In a two-dimensional case, the time consumed in image reconstruction is much saved, compared with that in the conventional GI. The experimental results show that the reconstructed images have good qualities. The convenience and flexibility of ghost imaging with Kronecker product of random matrices enable us to perform image encryption with permutation matrices in CGI. Only the correct permutation matrices and the matrix sequence are used in the reconstruction, the image can be decrypted. This method provides a new idea of encryption in ghost imaging, and ghost imaging with Kronecker product of random matrices may also have some application prospect in other image processing such as watermarking, image hiding, and so on.

## Acknowledgement

This work was financially supported by the National Natural Science Foundations of China under Grant Nos. 62105278 & 11674273.